\documentclass[aps,pre,preprint,superscriptaddress,showkeys]{revtex4-1}
 
\usepackage{graphicx}

\begin{document}

\preprint{}

\title{Criticality and the fractal structure of $-5/3$ turbulent cascades}

\author{Juan Luis Cabrera}
\email{juluisca@gmail.com}
\affiliation{Departamento de F\'isica Aplicada, ETSI Aeron\'autica y
  del Espacio, Universidad Polit\'ecnica de Madrid, Pza. Cardenal
  Cisneros 3, 28040, Madrid, Spain}
\affiliation{Laboratorio de Din\'amica Estoc\'astica, Centro de
  F\'isica, Instituto Venezolano de Investigaciones Cient\'ificas,
  Caracas 1020-A,  Venezuela}
\affiliation{Deeptikus Ltd., 10A Tasman Ave., Mount Albert, Auckland
  1025, New Zealand}
\author{Esther D. Guti\'errez} 
\affiliation{Facultad de Ciencias Naturales y Matem\'aticas,
  Escuela  Superior Polit\'ecnica del Litoral,    Km 30.5 V\'ia
  Perimetral,  Guayaquil, Ecuador}
\affiliation{Laboratorio de Din\'amica Estoc\'astica, Centro de
  F\'isica, Instituto Venezolano de Investigaciones Cient\'ificas,
  Caracas 1020-A,  Venezuela}
\affiliation{Deeptikus Ltd., 10A Tasman Ave., Mount Albert, Auckland  1025, New Zealand}
\author{Miguel Rodr\'iguez M\'arquez}
\affiliation{Departamento de Ciencias Naturales y Exactas,  Corporaci\'on Universidad de la Costa, Calle 58 No. 55-66,  Barranquilla, Colombia}
\affiliation{Deeptikus Ltd., 10A Tasman Ave., Mount Albert, Auckland  1025, New Zealand}
\date{\today}

\begin{abstract}
Here we show a procedure to generate an analytical
structure producing a cascade that scales as the energy spectrum in isotropic homogeneous turbulence.
We obtain a function that unveils a non-self-similar fractal at the origin of the cascade.
It reveals that the backbone underlying $-5/3$ cascades is formed
by deterministic nested polynomials with  parameters tuned in a Hopf bifurcation critical point. 
The cascade scaling is exactly obtainable (not by numerical simulations) from deterministic 
low dimensional nonlinear dynamics. Consequently, it should not be exclusive for fluids but also present in other complex phenomena. 
The scaling is obtainable both in deterministic and stochastic situations. 

\end{abstract}

\keywords{Cascade,  criticality,  fractals,  nonlinear,  stochastic,  turbulence,  maps,  complex.}

\maketitle


\section{Introduction}

Low dimensional dynamical systems  have played a central role in our understanding of the transition to turbulence in hydrodynamical systems 
\cite{eckmann,kadanoff} and provided a background to study turbulence \cite{vulpiani,frisch_1995,donough}. 
As L.P. Kadanoff once pointed out, the use of a simple system in hydrodynamics is similar to its use in critical phenomena in condensed matter physics, where it allows to understand complicated phase transitions because it shows qualitatively similar changes \cite{kadanoff}. The analysis of the simple system allows the extraction of universal features that are also found in the more complex problem.  
A remarkable example of such an approach has been using the logistic equation  \cite{May}, to explain the period doubling route to turbulence \cite{kadanoff,Feigenbaum,Libchaber}. 
Unfortunately, in spite of that insight and many others, turbulence still remains an open problem. 

In this paper we are concerned with the energy cascade observed in fully developed turbulence and characterized by a  $-5/3$ power law spectrum. 
Inspired by the idea of a cascade advanced previously by  Richardson \cite{richardson1926}, and based on a clever analysis, the scaling law was obtained by  Kolmogorov  \cite{kolmogorov}, Onsager \cite{onsager45} and others \cite{obukhov,heisenberg,vonWeizsacker}. 
Nevertheless, a detailed description of the underlying process that governs the flow of energy, through the different scales involved in a turbulent cascade, has yet to be found. We analyze a particular low dimensional system and obtain an analytical structure that describes a branching process yielding the $-5/3$ power law.  
This outcome is the result of an exact calculation, not using a numerical simulation. 

\section{The DRM}
We study a well known low dimensional dynamics that - as is the case of the logistic equation - 
is originated in population biology \cite{MaynardSmith68}.
 Let's consider  an stochastic version of the delayed regulation model (DRM),  given by the map on the unit interval, $x \in [0,1]$, 
\begin{eqnarray}  
x_{g+1} &=& r_g x_g (1 - x_{g-1}) \label{eq1}, 
\end{eqnarray}
\noindent
where $g=1,\dots,+\infty$,  is an iteration index we will call generation, 
$r_g = a\eta_g + b$, with $\eta_g$ some random perturbation indexed by $g$; $a$, the intensity and $b$ a bias parameter. 
For simplicity sake we assume the simplest case with $\eta_g$ uniformly distributed on  $[0,1]$, so that  $r_g \in [b, a+b]$.
For the case of zero delay and zero noise, the map (\ref{eq1}) is the deterministic logistic map.
Eq. (\ref{eq1}) is one of the simplest dynamical systems containing
nonlinearities, time delayed feedback and parametric random
perturbations.
The deterministic counterpart of Eq. (\ref{eq1})  has been widely
studied
\cite{aronson,pounder,morimoto} and its non-deterministic
version has been used to analyze stochastic extinction 
\cite{cabreraPhdthesis,cab1,cab2}, autonomous stochastic resonance \cite{cab3}
and noise-induced localization  \cite{cab4}.
Eq.  (\ref{eq1}) shows sustained on-off intermittency 
caused by destabilization of the origin and 
by destabilization of the fixed point  $1 - 1/<r_g>$, 
where  the intermittent temporal series contains bursts involving stochastic limit cycle oscillations. 
The DRM undergoes a Hopf bifurcation at $r_H=2$ \cite{morimoto, cab1,cab2}.

\section{Linealization of the DRM}

In the following, we will utilize noise to help with the calculations.
It will allow us to obtain simple expressions which in turn will lead us to the main results. 
Later, such results will be analyzed with and without noise.
The first step consist in the linearization of Eq. (\ref{eq1}), which is 
rewritten as, 
\begin{eqnarray}
x_{g+1} &=& r_g F(x_g, y_g) \\ \label{eq2}
y_{g+1}&=& G(x_g, y_g),
\end{eqnarray}
\noindent with $F(x, y) \equiv x (1 - y)$ and  $G(x, y) \equiv x$. $F$ and $G$
can be expanded around the fixed point $P \equiv (\alpha, \alpha)$, with
$\alpha \equiv 1 - \frac{1}{<r_g>} \equiv 1 -\frac{1}{\beta}  $, to obtain, 
\begin{eqnarray}
F(x, y) &=& F(\alpha, \alpha) + (x - \alpha) \frac{\partial F}{\partial x} \Big|_{P} + (y - \alpha) \frac{\partial F}{\partial y} \Big|_P + O(x^2, y^2)\\ \label{eq4}
G(x, y) &=& G(\alpha, \alpha) + (x - \alpha) \frac{\partial G}{\partial x} \Big|_P + (y - \alpha) \frac{\partial G}{\partial y} \Big|_P + O(x^2, y^2) , 
\end{eqnarray}
\noindent
i.e., 
\begin{eqnarray}
F(x, y) &=& \alpha^2 + (1 - \alpha) x - \alpha y \\ \label{eq5}
G(x, y) &=& x .
\end{eqnarray}
\noindent
Close to the fixed point $P$, Eq. (\ref{eq1}) can be approximated by its 
linear part: 
\begin{eqnarray}
  x_{g+1} &=&
              r_g \left( \alpha^2 +  \frac{x_g}{\beta} - \alpha y_g \right)\\ \label{eq6}
y_{g+1} &=& x_g   
\end{eqnarray}
\noindent
And it is better to rewrite it as, 
\begin{equation}  
\left(
\begin{array}{c}
    x \\
    y 
\end{array}
\right)_{g+1}
    = 
\left(
   \begin{array}{cc}
    r_g /\beta & -\alpha r_g \\ 
    1               & 0 
   \end{array} 
\right)
\left(
   \begin{array}{c}
    x \\
    y 
   \end{array}
\right)_{g} + 
\left(
   \begin{array}{c}
    \alpha^2 r_g \\ 
    0
   \end{array}
 \right)   \label{lin} .
\end{equation}  
\noindent So we can define the evolution matrix: 
\begin{equation}  
\mathbf{  A_g }\equiv  
\left(
   \begin{array}{cc}
    r_g /\beta & - \alpha r_g\\ 
    1               & 0 
   \end{array} 
 \right), \label{A}
\end{equation}   
\noindent 
and the bias vector:
\begin{equation}  
\vec{B}_g  \equiv  
\left(
   \begin{array}{c}
    \alpha^2 r_g \\ 
    0 
   \end{array} 
 \right) . \label{B2}
\end{equation}
\noindent Then, Eq. (\ref{lin}) can be compacted as,
\begin{equation}  
  \vec{X}_{g+1} = \mathbf{ A_g }\vec{ X}_g  \,+ \vec{ B}_g. \label{eq7}
\end{equation}  

\section{Expanding  $\vec{ X_g }$}

It is useful to express $\vec{ X_g }$ in terms of the initial state $\vec{ X_0 }$. 
So, let's note that  $\vec{X}_{g+1}$ can be patiently expanded as follows,
\begin{eqnarray}
\vec{X}_{g+1} &=& \mathbf{ A_g }\vec{ X_g } + \vec{B}_g   \nonumber \\
&=& \mathbf{ A_g ( A_{g-1} }\vec{X}_{g-1}  + \vec{B}_{g-1} ) + \vec{B}_g  \nonumber \\
&=& \mathbf{ A_g ( A_{g-1} ( A_{g-2} }\vec{X}_{g-2}  + \vec{B}_{g-2} ) + \vec{B}_{g-1} ) + \vec{B}_g  \nonumber \\
 &\;\;\vdots &  \nonumber \\
&=& \mathbf{ A_g A_{g-1} A_{g-2} A_{g-3} ...  A_1A_0 }\vec{ X}_0    \nonumber \\
&+& \mathbf{ A_g A_{g-1} A_{g-2}  A_{g-3} ...  A_1 }\vec{ B}_0   \nonumber \\
 &\;\;\vdots &  \nonumber \\
&+& \mathbf{ A_g A_{g-1} A_{g-2} }\vec{ B}_{g-3}   \nonumber \\
&+& \mathbf{ A_g A_{g-1} }\vec{ B}_{g-2}   \nonumber \\
&+& \mathbf{ A_{g} }\vec{ B}_{g-1}  \nonumber \\
&+& \vec{ B}_g  
\end{eqnarray}
\noindent
So, $\vec{X}_{g+1}$ can be written as, 
\begin{eqnarray}
\vec{X}_{g+1} &=& \prod_{j=0}^{g} \mathbf{  A_{g-j} }\vec{ X_0 } \nonumber \\
&+&  \prod_{j=0}^{g-1} \mathbf{ A_{g-j} }\vec{ B_0 }  \nonumber \\
&+& \prod_{j=0}^{g-2} \mathbf{ A_{g-j} }\vec{ B_1 }  \nonumber \\
&+& \prod_{j=0}^{g-3} \mathbf{ A_{g-j} }\vec{ B_2 }  \nonumber \\
&\;\;\vdots &  \nonumber \\
&+& \prod_{j=0}^{g-(g-2+1) = 1} \mathbf{ A_{g-j} }\vec{ B}_{g-2} \nonumber \\
&+& \prod_{j=0}^{g-(g-1+1) = 0} \mathbf{ A_{g-j} }\vec{ B}_{g-1}   \nonumber \\
&+& \vec{ B}_{g}  \nonumber \\
&=& \prod_{j=0}^{g} \mathbf{ A_{g-j} }\vec{ X_0 } \nonumber \\
&+&  \sum_{i=1}^{g+1} \prod_{j=0}^{g-i} \mathbf{ A_{g-j} }\vec{ B}_{i-1}  .
\end{eqnarray}
\noindent 
Where we are using the definition,
\begin{eqnarray}
	\prod_{j=0}^{g-i} \mathbf{A_{g-j} } &\equiv& 1 \; \text{if} \;  i=g+1 .
\end{eqnarray}
\noindent
Now, changing  variables, $g \longrightarrow g' - 1$, we obtain,
\begin{eqnarray}
\vec{ X_{g}} &=& \prod_{j=0}^{g-1} \mathbf{ A_{g-j-1} }\vec{ X_0 } +  \sum_{i=1}^{g} \prod_{j=0}^{g-i-1} \mathbf{ A_{g-j-1} }\vec{ B}_{i-1}  ,
\end{eqnarray}
\noindent
where we have omitted the prime and used,
\begin{eqnarray}
\prod_{j=0}^{g-i-1} \mathbf{ A_{g-j-1} }&\equiv& \mathbf{ 1} \;
                                                 \text{if} \; i=g .
\end{eqnarray}
\noindent 
Thus,  we can define, 
\begin{eqnarray}
\mathbf{ P_i }& \equiv & \prod_{j=0}^{g-i-1} \mathbf{ A_{g-j-1} } \;\;\;\;\;\;\;\; \text{if} \; i \neq g  ,\\
\mathbf{ P_i }& \equiv & \mathbf{ 1 }\; \;\;\;\;\;\;\;\;\;\;\;\;\;\;\;\;\;\;\;\;\;\;\; \text{if} \;  i=g  ,\nonumber \\
\mathbf{ P_g }& \equiv &  \prod_{j=0}^{g-1} \mathbf{ A_{j}},
\end{eqnarray}
\noindent
to write $\vec{X}_{g}$ in a compact form, 
\begin{eqnarray}
\vec{X}_{g} &=& \mathbf{ P_g }\vec{X}_0     +  \sum_{i=1}^{g} \mathbf{ P_i }\vec{B}_{f(i)}  , \label{l1}
\end{eqnarray}
\noindent with $f(i) \equiv i-1$.

\section{Simplifying $ \vec{X}_{g} = \mathbf{ P_g} \vec{X}_0     +  \sum_{i=1}^{g} \mathbf{ P_i }\vec{B}_{f(i)}   $}
For convenience, the following calculations are carried out on the complex plane. 
Here we want to simplify the expression (\ref{l1}) as much as possible. 
So, let be $\vec{Y}_{g}$ a state at  generation $g$, obtained with a
realization $r_{g}^{\prime} \neq r_{g}$. 
Calculating the dot product of $\vec{X}_{g}$  and $\vec{Y}_{g}$ gives, 
\begin{eqnarray}
 \vec{X}_{g} \cdot  \vec{Y}_{g}  &=& \mathbf{  P_g} \vec{X}_0  \cdot  \vec{Y}_{g} +   \sum_{i=1}^{g} \mathbf{ P_i }\vec{B}_{f(i)}  \cdot  \vec{Y}_{g}  . \label{l2}
\end{eqnarray}
\noindent The scalar quantity, $\eta_g  \equiv  \vec{X}_{g}  \cdot  \vec{Y}_{g} $, is the value for the projection of $\vec{ X}_{g}$ onto  $\vec{Y}_{g}$. 
It is a random variable evaluated at generation $g$, that takes values on the interval $[0, 1]$.
Also, let's note that applying the operator $\mathbf{ P_g}$ on the initial
condition $\vec{ X}_0$, produces a new state at  generation $g$, 
say $\vec{ \mu}_g $. Thus, the dot product between $\mathbf{ P_g} \vec{ X}_0  = \vec{ \mu}_g $ 
and $\vec{Y}_{g} $
yields also a scalar random variable at generation $g$, 
say $0 \leq \psi_g \leq 1 $ , i.e., 
\begin{eqnarray}
 \mathbf{P_g} \vec{X}_{0} \cdot  \vec{Y}_{g} =   \vec{\mu}_g \cdot  \vec{Y}_{g}   & \equiv &  \psi_g  \label{mu}
\end{eqnarray}
\noindent With these considerations in mind we can rewrite Eq. (\ref{l2}) as,
\begin{eqnarray}
\eta_g &=& \psi_g     +    \sum_{i=1}^{g} \mathbf{ P_i }\vec{B}_{f(i)}  \cdot  \vec{Y}_{g}\nonumber \\  \label{l3} 
& = &  \psi_g    \left(  1 +  \psi_{g}^{-1}   \sum_{i=1}^{g} \mathbf{  P_i }  \vec{ B}_{f(i)}  \cdot  \vec{Y}_{g}  \right) .
\end{eqnarray}
\noindent
After rearranging terms, it results in,
\begin{eqnarray}
 \psi_{g}   \left(  \frac{\eta_g}{ \psi_g } - 1 \right) &=&
                                                           \sum_{i=1}^{g} \mathbf{  P_i}   \vec{ B}_{f(i)}  \cdot  \vec{Y}_{g}
\end{eqnarray}
\noindent or 
\begin{eqnarray}
 \left(  \frac{\eta_g}{ \psi_g } - 1 \right) \vec{\mu_g }   \cdot  \vec{Y}_{g} &=& \sum_{i=1}^{g}  \mathbf{ P_i}   \vec{ B}_{f(i)}   \cdot  \vec{Y}_{g} \label{etamu},
\end{eqnarray}
\noindent where we have used Eq. (\ref{mu}). This expression is the same
as,
\begin{eqnarray}
 \Bigg\{ \left(  \frac{\eta_g}{ \psi_g } - 1 \right)
  \vec{ \mu}_g  -  \sum_{i=1}^{g}
                           \mathbf{ P_i }  \vec{ B}_{f(i)}  
                            \Bigg\} \cdot  \vec{Y}_{g}&=& 0\label{parentesis}.
\end{eqnarray}

\noindent Now, at a given step $g$, this equation has two possible consequences: i) either the vectors,  
$ \left( \left(  \frac{\eta_g}{ \psi_g } - 1 \right)
  \vec{ \mu_g }  -  \sum_{i=1}^{g}  \mathbf{ P_i }  \vec{B}_{f(i)} 
\right) $ and $\vec{Y}_{g}$, are orthogonal vectors, in which case its dot product would equal zero, or ii) the following equality is satisfied, 
\begin{eqnarray}
 \left(  \frac{\eta_g}{ \psi_g } - 1 \right)  \vec{ \mu}_g   &=&
                                                                 \sum_{i=1}^{g}
                                                                 \mathbf{
                                                                 P_i }  
                                                                 \vec{
                                                                 B}_{f(i)}  \label{equal1}, 
\end{eqnarray} 
In what follows, we assume that condition ii) predominates, i.e., that the involved random dynamics reduces the probability of i) to a minimum, 
such that most of the evolution of the system can be better characterized by condition ii) at least in 
$O(
\{ \left(  \frac{\eta_g}{ \psi_g } - 1 \right)
  \vec{ \mu}_g  -  \sum_{i=1}^{g}
                           \mathbf{ P_i }  \vec{ B}_{f(i)}
                           \}   \cdot \vec{Y}_{g}
)$. 
That being said, we must also impose  $b>1$, to avoid the case $r_g < 1$ if $\eta_g=0$, which would take the system state to the zero fixed point, a situation where we can't rule out  $\vec{Y}_{g}=0$, at least momentarily. 
With these thoughts in mind, and using the definition of $\vec{\mu}_g $, condition ii) turns out as,
\begin{eqnarray}
 \left(  \frac{\eta_g}{ \psi_g } - 1 \right)  \mathbf{ P_g }\vec{ X}_0    &=&   \sum_{i=1}^{g}  \mathbf{ P_i }  \vec{ B}_{f(i)}  \label{etamu2}, 
\end{eqnarray}
\noindent
a relation that we can introduce into Eq. (\ref{l1}), to obtain a shorter
expression for the time evolution of the system in terms of the product $\mathbf{ P_g }$, 
\begin{eqnarray}
\vec{ X}_{g} &=& \mathbf{ P_g }\vec{ X}_0     +   \left(  \frac{\eta_g}{ \psi_g } - 1 \right)  \mathbf{  P_g }\vec{ X}_0  ,\nonumber \\
&=&     \frac{\eta_g}{ \psi_g }    \mathbf{ P_g } \vec{ X}_0  \nonumber \\
&=&     \gamma_g  \mathbf{   P_g } \vec{ X}_0  \nonumber \\
&=&      \mathbf{  P_{g}^{\prime} } \vec{ X}_0  . 
\end{eqnarray}
\noindent
Here, $\gamma_g \equiv \frac{\eta_g}{\psi_g}$, is a  new random number and $
\mathbf{ P_{g}^{\prime} }\equiv \gamma_g   \mathbf{ P_g }$ is the original product modulated by $\gamma_g$. 
To know the range of values that $\gamma_g$ takes on,  let's  consider the inequality:
\begin{eqnarray}
0 \leq  \mathbf{  P_g } \vec{ X}_0 \cdot  \vec{Y}_{g}   <
   \mathbf{  P_g } \vec{ X}_0 \cdot  \vec{Y}_{g}     +
   \sum_{i=1}^{g} \mathbf{  P_i } \vec{ B}_{f(i)} \cdot  \vec{Y}_{g} 
    &=&       \vec{ X}_{g} \cdot  \vec{Y}_{g}  \leq 1   \label{ine1}
\end{eqnarray}
\noindent or
\begin{eqnarray}
0 \leq      \mathbf{  P_g } \vec{ X}_0  \cdot  \vec{Y}_{g}  <  \vec{ X}_{g} \cdot  \vec{Y}_{g}   \leq  1 ,  \label{ine2}
\end{eqnarray}
\noindent i.e.,
\begin{eqnarray}
0 \leq  \psi_g  <    \eta_g  \leq 1 . \label{ine3}
\end{eqnarray}
\noindent
Then, 
\begin{eqnarray}
1 < \gamma_g  < \infty .
\label{ine4}
\end{eqnarray}
\noindent
Consequently, $\gamma_g$ is an unbounded random variable larger than $1$, i.e., 
applying $\gamma_g$ on $\mathbf{  P_g }$ has an 
amplifying effect.

\section{Calculating the norm of $ \vec{ X}_{g}$}
At this point, we would like to know about the behavior of the norm of $ \vec{ X}_{g}$. 
Such a quantity can be determined calculating the following inner product,
\begin{eqnarray}
\parallel  \vec{X}_g \parallel =  [\vec{ X}_{g}^{*}  \vec{ X}_g  ]^{1/2}  =
  [  \vec{  X}_0 \mathbf{  P_{g}^{' \dag} P_{g}^{\prime} } \vec{ X}_0  ]^{1/2}  \label{inner1}, 
\end{eqnarray}
\noindent
given that $\mathbf{ P_g^{\prime} }$ is not a self-adjoint operator. Here $\mathbf{  P_{g}^{' \dag} }$  is the Hermitian conjugate of the operator $\mathbf{  P_{g}^{\prime} }$, i.e., 
the adjoint matrix in our case. Then, the problem of calculating the norm translates to the calculation of 
$\mathbf{ P_{g}^{' \dag} P_{g}^{\prime} } $. 
To evaluate such a product we must write it in terms of the time
dependent random matrix $\mathbf{ A_g }$,  
\begin{eqnarray}
\mathbf{ P_{g}^{' \dag} P_{g}^{\prime} } &=&  \gamma_g^{*} \mathbf{ P_{g}^{\dag} } \gamma_g \mathbf{  P_{g} } = \gamma_g^{*} \gamma_g  \mathbf{ P_{g}^{\dag}  P_{g} },
\end{eqnarray}
\noindent where
\begin{eqnarray}\label{40}
 \mathbf{ M_{g} } \equiv \mathbf{ P_{g}^{\dag}  P_{g} } & = & [ \mathbf{A_{g-1}A_{g-2} ... A_{1}A_{0} ]^{\dag}} 
 \mathbf{ A_{g-1}A_{g-2} ... A_{1}A_{0} } = \nonumber \\
& = &   \mathbf{ A_{0}^{\dag} A_{1}^{\dag} ... A_{g-2}^{\dag} A_{g-1}^{\dag} A_{g-1}A_{g-2} ... A_{1}A_{0} } = \nonumber \\
& = &  \mathbf{ A_{0}^{\dag} [ A_{1}^{\dag} ... [ A_{g-2}^{\dag} [ A_{g-1}^{\dag} A_{g-1} ] A_{g-2} ] ... A_{1}  ]  A_{0} }, 
\end{eqnarray}
\noindent 
and
\begin{equation}  
\mathbf{A_g^{\dag} }= 
\left(
   \begin{array}{cc}
    r_g /\beta & 1 \\ 
    -r_g \alpha              & 0 
   \end{array} 
\right) \label{A+}
\end{equation}

\section{Eigenvalues of $p_1$}

A straightforward evaluation of $ \mathbf{ M_{g} } $ doesn't seem plausible. 
However, we can address our attention on its step by step evolution.
In particular, in the equation for $ \mathbf{ M_{g} }$, the core or first 
product is given by:
\begin{equation}  
p_1\equiv \mathbf{A}_{g-1}^{\dagger }\mathbf{A}_{g-1}=\left(
\begin{array}{cc}
\frac{r_{1}^2+\beta^2}{\beta^2} & -\frac{r_{1}^2}{\beta} \alpha \\ 
-\frac{r_{1}^2}{\beta}a & \alpha^2r_{1}^2
\end{array}
\right) , \label{p1}
\end{equation}  
\noindent where we are using the notation $ r_i \equiv r_{g-i}$. It
should be noted that this matrix has eigenvalues,
\begin{eqnarray}
\lambda}_{\pm}^{p_{1} &=& \frac 1{2\beta^2}\left(  \left( 1+\alpha
                          ^2r_1^2\right) \beta ^2+r_1^2  
                          \pm \sqrt{
                          \left(  \left( 1+\alpha ^2r_1^2\right) \beta ^2+r_1^2\right) ^2-        \left( 2 \alpha \beta^2r_{1}\right) ^2 
                          } \right) .
\end{eqnarray}
\noindent 
So, the eigenvalues of the product $p_1$  can be expressed as, 
\begin{eqnarray}
\lambda}_{\pm}^{p_{1}  &=&\frac 1{2 \beta^2}\left( \Lambda_1 \pm
                           \sqrt{\Lambda_1^2- \Upsilon_1^2} \right) .
\end{eqnarray}
\noindent where, $  \Lambda_1 \equiv \left( 1+\beta ^2\alpha ^2\right) { r_1}^2+\beta ^2 , \label{A1}$  and $\Upsilon_1\equiv 2 \beta^2 \alpha r_{1}$. 
This process is repeated to calculate the next product, 
$p_2\equiv \mathbf{A}_{g-2}^{\dag}  p_1\mathbf{A}_{g-2}$, and
similarly, $p_3$, and so on. A detailed account of these calculations is
provided in the appendixes A-D. From there, it is clear that 
the eigenvalues for the first five products can be expressed as: 
\begin{eqnarray}
\lambda _{\pm}^{p_2}&=&\frac 1{2\beta^4}\left( \Lambda_2 \pm \sqrt{\Lambda_2^2-\Upsilon_2^2}\right) \\
\lambda _{\pm}^{p_3}&=&\frac 1{2\beta^6}\left( \Lambda_3 \pm \sqrt{\Lambda_3^2-\Upsilon_3^2}\right) \\
    \lambda _{\pm}^{p_4} &=&\frac 1{2\beta^8}\left( \Lambda_4 \pm \sqrt{\Lambda_4^2-\Upsilon_4^2}\right) \\
\lambda _{\pm}^{p_5}&=&\frac 1{2\beta^{10}}\left( \Lambda_5 \pm \sqrt{\Lambda_5^2-\Upsilon_5^2} \right)
\end{eqnarray}

\noindent 
where the definitions for $\Lambda_i$ and $\Upsilon_i$ are given in the appendixes. 

\begin{figure}[t]
\centering
\includegraphics[width=1\textwidth]{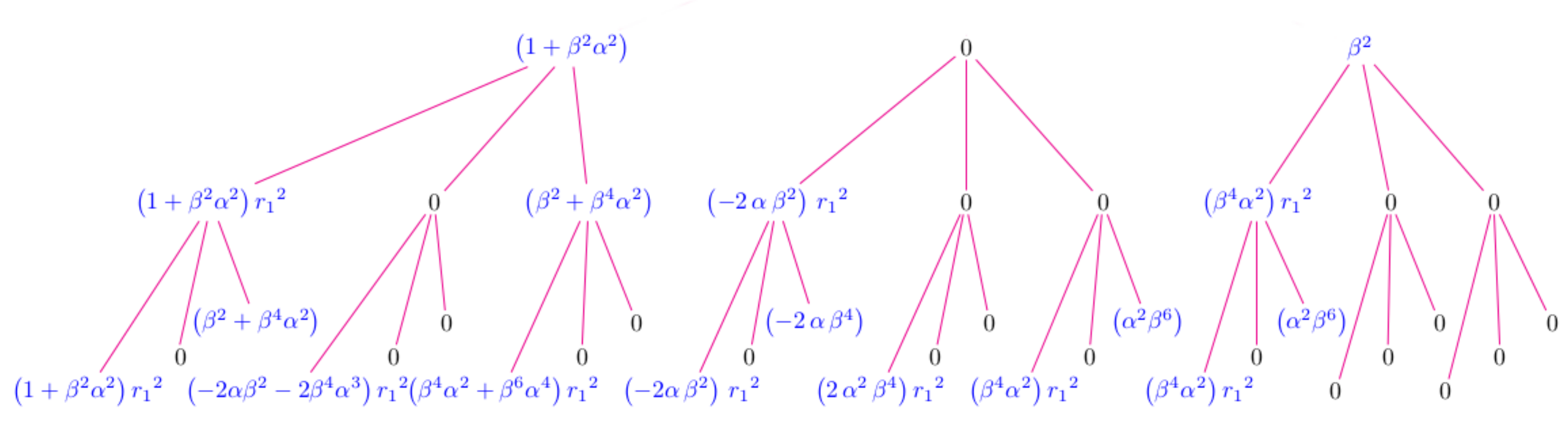}
\caption{Tree representation for the generation of  the polynomial
  coefficients after two iterations of rule 
  (\ref{rule0}, \ref{rule1}), starting from $\Lambda_1$. Bottom coefficients
  are those forming $\Lambda_3$ given by Eq. (\ref{A3}).} 
\label{fig1}
\end{figure}

\section{Nested structures}

The previous section allows us to infer how the eigenvalues evolve with each generation $g$. 
In particular, the eigenvalues for the products, $p_1, p_2, ... , p_N$, can be written as, 
\begin{eqnarray} \label{eigenvalues}
\lambda _{\pm}^{p_g}=\frac 1{2\beta^{2g}}\left( \Lambda_g \pm \sqrt{\Lambda_g^2-\Upsilon_g^2}%
  \right), 
\end{eqnarray}
\noindent
where terms $\Lambda_g$ and $\Upsilon_g$ are polynomials in the noise terms, i.e.,
random polynomials. In particular, 
$\Upsilon_g$, can be easily written in compact form as, 
\begin{eqnarray}\label{B}
\Upsilon_g = 2 \beta^{2g} \alpha^{g} \prod_{i=1}^{g} r_{i}.
\end{eqnarray}
\noindent
However, a closed form for $\Lambda_g$ can not be obtained so easily.
It is so because $\Lambda_g$ follows a complex nested structure as is 
illustrated in the appendixes.
One must note that $\Lambda_g$ is a polynomial in $r_g$ of order two, which coefficients 
are polynomials of order two in $r_{g-1}$,  which coefficients  
are polynomials of order two in $r_{g-2}$,  and so on. 
So, let's say that $C( r_{g-1} , \dots , r_{1})$ is a function that depends
on the noise terms $ r_{g-1} , \dots , r_{1}$, and that such a
function is a coefficient of a noise term of order $i$, i.e., that given 
a second order polynomial describing $\Lambda_g$  it has terms $C_{i}(r_{g-1}, \dots , r_{1})r^i$. Now,
depending on what $\Lambda_g$ are we dealing with, we shall distinguish each
of these functional coefficients. Consequently, it is convenient
indexing also the $C$'s to make such a distinction, so writing these terms as 
 $C_{g,i}(r_{g-1}, \dots , r_{1})r_g^i$. Here, we have
also indexed $r_g$, because that noise's value  is exactly the one
at step $g$. Accordingly, $\Lambda_g$ can be conveyed to, 
\begin{eqnarray} \label{An}
\Lambda_g & = & \sum_{i=0}^2 C_{g,i} ( r_{g-1}, ... , r_{1} ) r_g^i 
\end{eqnarray}
\noindent with $  C_{1,i} \equiv  C_{g,i} ( r_{g-1}, ... , r_1 ) |_{g=1} $,  constant. 
Note that we have not indexed the coefficients of the nested levels because there is no need for that, 
we try to keep the notation simple, as far as possible.  
Now, let's reproduce the full structure as follows.  
In general, if we know $\Lambda_1$ and $\Lambda_2$, we can obtain $\Lambda_g$ given that, in the generation $g-1$, 
a given term,
\begin{eqnarray}
\left\{ C_{2} r_{i}^2 + C_{1} r_{i} + C_{0}   \right\} r_{i+1}^{k} 
  \,\,\,\,\, \text{with} \,\,\,\,\, k=0,1,2 , \label{rule0}
\end{eqnarray}
\noindent generates the polynomial, 
\begin{eqnarray}
\left\{  
\begin{array}{c}
\left[ C_{2} r_{i}^2 + C_{1} \chi_1 r_{i} + \beta^2 C_{2} \right] r_{i+1}^{2} \\
+ \left[  -k \alpha \beta^2 C_2 r_i^2 + C_1 \chi_2  r_i + C_1 \chi_3   \right] r_{i+1} \\
+ \left[ \beta^2 \alpha^2 C_0 r_i^2 + C_1 \chi_4 r_i + C_1 \chi_5 \right]  
\end{array}
\right\} r_{i+2}^{k} \label{rule1},
\end{eqnarray}
\noindent in the next generation, $g$. $\chi_1, \chi_2, ... , \chi_5$ 
are unknowns, here included to stress the polynomial nested structure, but multiplied by  $C_1$,  that in the present
situation vanishes.

\section{Generating nested estructures: an example}

The preceding procedure is better illustrated with an 
example:  finding $\Lambda_3$ from $\Lambda_2$, which is given by
Eq. (\ref{A2}). 
\noindent  The coefficient for the quadratic term is, 
$ \left[ (1+ \alpha^2\beta^2) r_{1}^2 + \beta^2 (1+\alpha^2\beta^2)  \right], $
therefore, $C_2=1 + \alpha^2\beta^2$, $C_1=0$ and $C_0 = \beta^2 (1+ \alpha^2 \beta^2)$; and the 
quadratic term in $r_3$ is generated as,
\begin{eqnarray}
\left\{  
\begin{array}{c}
\left[ (1+\alpha^2\beta^2) r_{1}^2 + \beta^2 (1+\alpha^2\beta^2) \right] r_{2}^{2} \\
+ \left[  -2 \alpha \beta^2 (1+\alpha^2\beta^2) r_1^2  \right] r_{2} \\
+ \left[ \beta^4 \alpha^2 (1+\alpha^2\beta^2) r_1^2 \right]  
\end{array}
\right\} r_{3}^{2} . \label{r32}
\end{eqnarray}
\noindent Now, the lineal term in $\Lambda_2$ is $-2\alpha \beta^2 r_1^2$,
therefore in this case, $C_2= -2 \alpha \beta^2$ and $C_1=C_0=0$.
Then the linear term in $r_3$ is,
\begin{eqnarray}
\left\{  
\begin{array}{c}
\left[ (-2 \alpha \beta^2) r_{1}^2 + (-2 \alpha \beta^4) \right] r_{2}^{2} \\
+ \left[  2 \alpha^2 \beta^4 r_1^2  \right] r_{2} 
\end{array}
\right\} r_{3} .\label{r31}
\end{eqnarray}
Finally, the coefficient of the independent term in $\Lambda_2$ is $
(\alpha^2\beta^4) r_1^2$. Then $C_2= \alpha^2\beta^4$ and $C_1=C_0=0$,
so we obtain, 
\begin{eqnarray}
\left\{  
\begin{array}{c}
\left[ (\alpha^2 \beta^4) r_{1}^2 + \alpha^2 \beta^6) \right]
  r_{2}^{2} 
\end{array}
\right\} r_{3}^0 . \label{r30}
\end{eqnarray}
\noindent
Adding Eq. (\ref{r32}), (\ref{r31}) and (\ref{r30}) results in $\Lambda_3$.

\begin{figure}[t]
\centering
\includegraphics[]{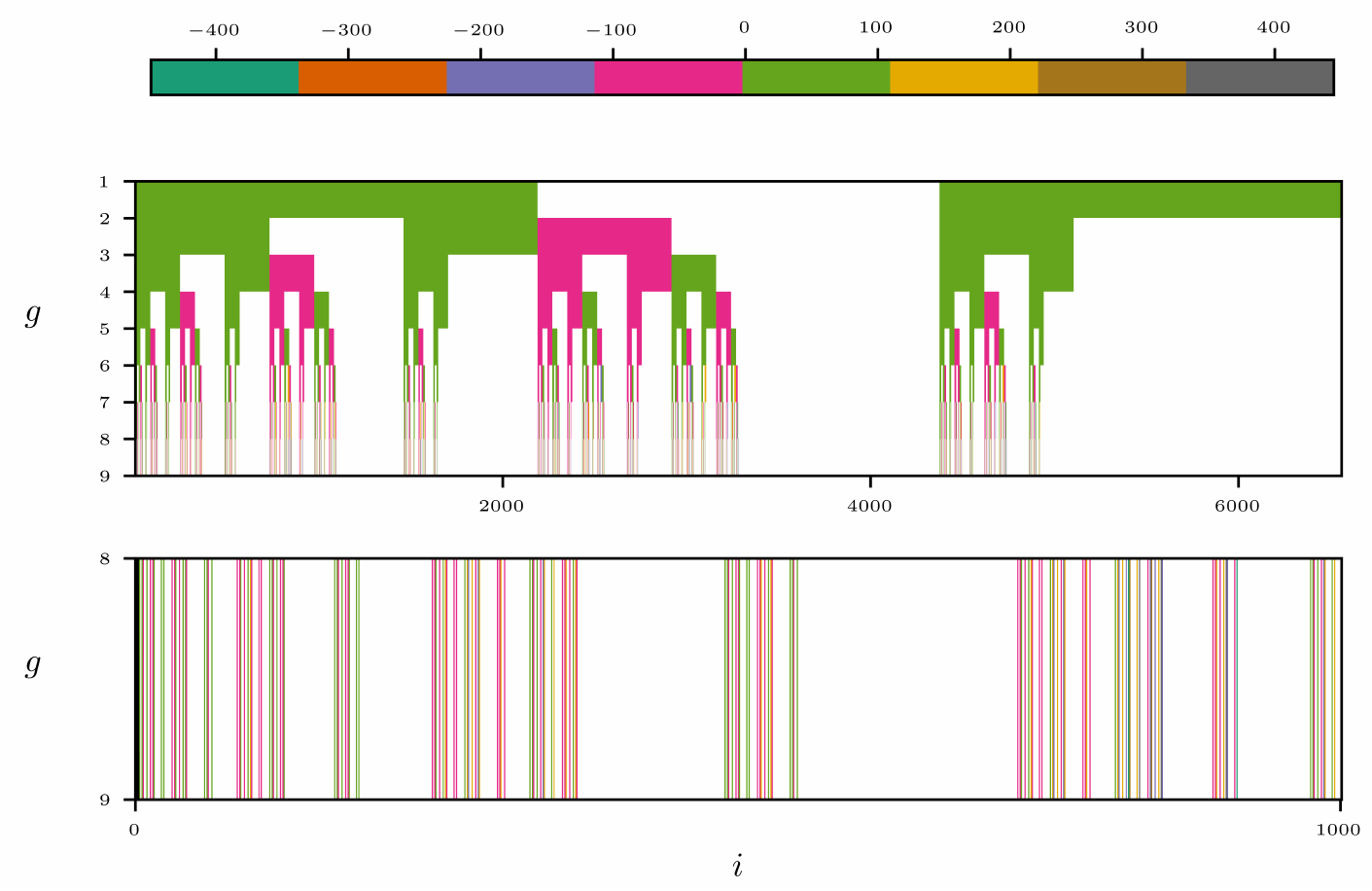}
\caption{(Top) Analytical deterministic cascade describing the evolution of the nested polynomial coefficients $C_2$, $C_1$ and $C_0$ (from left
  to right) after
  $n=8$ iterations. Colors
  represent the coefficient's  intensities $a_i$.
  The iteration process starts with the $\Lambda_1$ coefficients, each one
  laying on $\frac{1}{3}$ segments on the $x-$axis 
  which hosts $3^8$ subdivisions.
  (Bottom) Zoom in the interval $[0,1000]$ showing fine
  structure. 
  Parameter values are $r=1$, $a=0.085$ and $b=1.9375$.
} \label{fig2}
\end{figure}

\section{Determining the leading term in $\lambda$ }

The eigenvalues of
$\mathbf{ M_{g}}$ are given by Eq. (\ref{eigenvalues})
with the term $\Upsilon_g$ growing as a
power of the noise term, i.e., $\Upsilon_g \sim r^g$. 
Meanwhile, the  equations (\ref{A1}), (\ref{A2}), (\ref{A3}), (\ref{A4}) and
(\ref{A5}) indicate that the $\Lambda_g$'s grow as a sum of powers in the noise, i.e.,
\begin{eqnarray}
  \Lambda_{1}  &\sim & r^2 \nonumber\\
  \Lambda_{2}  &\sim & r^4 + r^ 3 + r^2 \nonumber\\
  \Lambda_{3}  &\sim & r^6 + r^5 + r^4 + r^ 3 + r^2 \\
  \Lambda_{4}  &\sim & r^8 + r^7 + r^6 + r^5 + r^4 \nonumber\\
  \Lambda_{5}  &\sim & r^{10} + \dots +  r^4 \nonumber.
\end{eqnarray}
\noindent
So, $\Lambda_g > \Upsilon_g$ and $\Lambda_g$ will predominate for $g$
sufficiently large, i.e.,  $\frac{\Upsilon_g}{\Lambda_g} \rightarrow 0$, thus
Eq. (\ref{eigenvalues}) yields,
\begin{eqnarray}\label{aprox}
\lambda _{+}^{p_g} \sim \Lambda_g +
  O\Bigg(\frac{\Upsilon_g}{\Lambda_g} \Bigg)\\
\lambda _{-}^{p_g}\sim  0  + O\Bigg(\frac{\Upsilon_g}{\Lambda_g}\Bigg)\nonumber
\end{eqnarray}

\noindent
Let's  unpack  Eq. (\ref{inner1}) to determine the implications of
this approximation on the norm $|| \vec{X}_g||$,
\begin{eqnarray}\label{unpack} 
\parallel  \vec{X}_g \parallel &=&   ( \gamma_g^{*} \gamma_g    \vec{X}_{0}^{*} \mathbf{  M_{g} } \vec{X}_0    )^{1/2}  \\
   &=&  \gamma_g\Bigg[   (  x_0  y_0 ) \left(
   \begin{array}{cc}
    \lambda_+ & 0 \\ 
    0               & \lambda_-
   \end{array} 
                      \right)
\left(
   \begin{array}{cc}
    x_0  \\ 
    y_0 
   \end{array} 
  \right)
  \Bigg]^{\frac{1}{2}} . \nonumber
\end{eqnarray}
\noindent
Combining this equation with the Eqs. (\ref{aprox}) we find that $\Lambda_g$ leads the behavior of the norm,
\begin{eqnarray}\label{dominance} 
\parallel  \vec{X}_g \parallel &\sim& \frac{\gamma_g}{\beta^g}
                               \sqrt{ \Lambda_{g}} x_0  \label{r33}.
\end{eqnarray}

\begin{figure}[htb!]
\centering
  \includegraphics[]{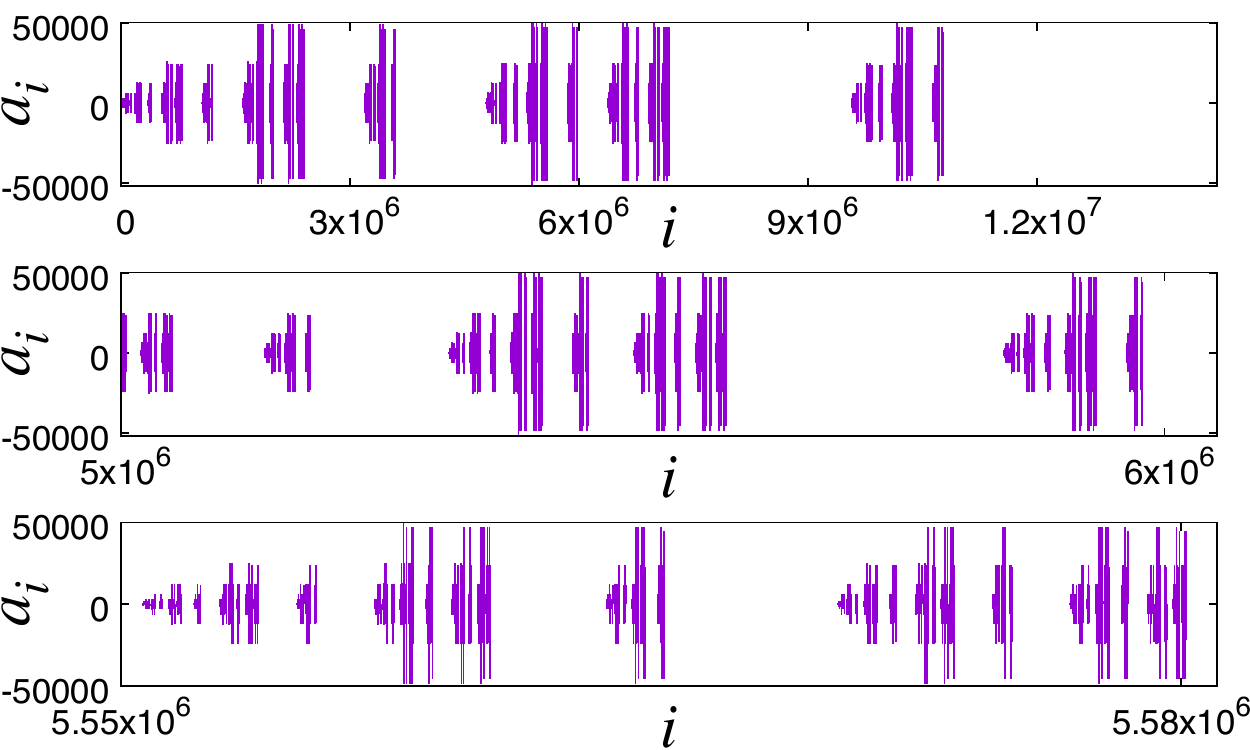}
\caption{Non-self-similar structure of the analytical deterministic cascade revealed
after successive enlargements. Image obtained with $g=15$ generations and parameter values as in figure 
\ref{fig2}. The number of coefficients calculated is
$3^{15}=14348907$. Points are represented by impulses and zero ones
have been extracted. 
} \label{fig3}
\end{figure}

\section{A hidden fractal structure}

\begin{figure}[htb!]
\centering
\includegraphics[]{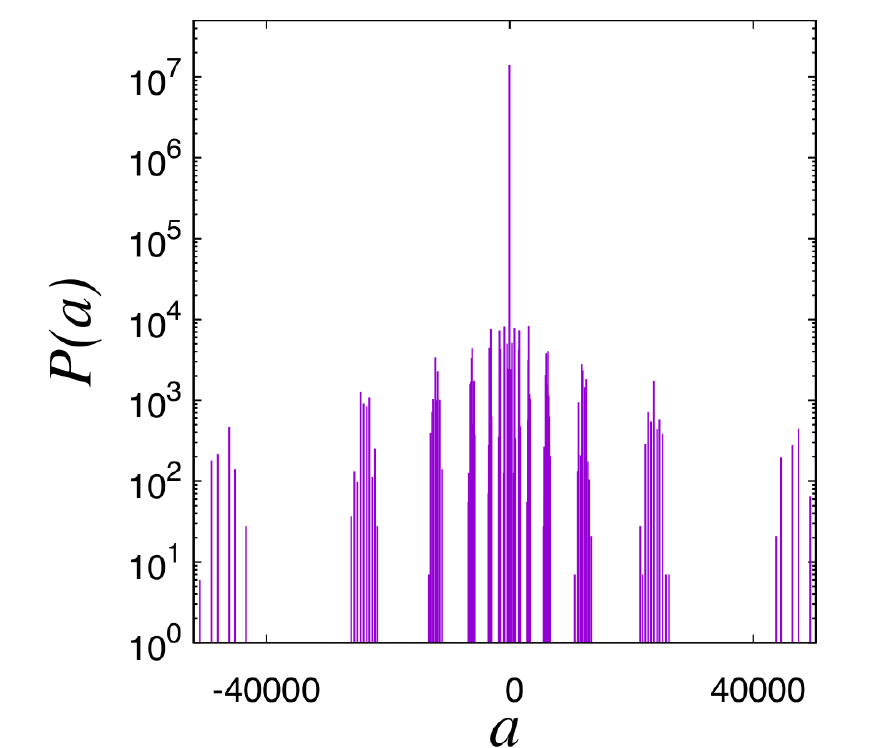}
\caption{ Semi-log plot of the intensity distribution from the
  deterministic cascade obtained  after $15$ generations. Same
  parameter values as in figure \ref{fig2}. }\label{fig4}
\end{figure}

The hidden structure of the $\Lambda_g$'s is better understood using a graphical representation. 
Figure \ref{fig1}  shows the branching process obtained after three iterations of the rule (\ref{rule0}, \ref{rule1}).
The terms displayed at the bottom are the coefficients of $\Lambda_3$
(see Appendix B).
This representation reveals a cascading process at the backbone of the eigenvalues time evolution.
To get a deeper insight into such a structure we will focus exclusively on the evolution ofa the coefficients  
described in figure \ref{fig1}, turning off any random modulation. It is achieved,  
arbitrarily considering $r_g=r$, a constant, conveniently established as $r=1$, $\forall g$. 
Therefore, now the magnitude of the coefficients depends only on the values of $a$ and $b$ 
 which will be chosen such that $(a+b)$ is close to the Hopf bifurcation parameter value $r_H=2$. 
This procedure allows us to separate the contribution of the coefficients exclusively
 as in figure \ref{fig1}.  
 Under these conditions the cascade obtained 
is depicted in figure \ref{fig2}. 
The method used to obtain this figure is as follows: 
the coefficient values in the initial polynomial $\Lambda_1$, 
 i.e., $C_2$, $C_1$ and $C_0$, define  branch intensities, $a_i$, $i = 1, 2, 3$, on the  $N_1=3^1$ subintervals of the unit interval:   
$I_2 \equiv [0,\frac{1}{3})$, $I_{1} \equiv [\frac{1}{3},
\frac{2}{3})$ and $I_{0} \equiv [\frac{2}{3},1]$, respectively.
After one iteration, each subset is divided by a factor of  $3$, and
the newly generated coefficients - obtained using rule (\ref{rule0}, \ref{rule1}) - shall update the $a_i$'s on 
the new $N_2=3^2$ subintervals.
After each new generation,
the unit interval is divided by a factor of $3$ such that, after
$g$ iterations, the value of the coefficients can be represented as intensities on $N_g=3^g$ subintervals,
 given by rule (\ref{rule0}, \ref{rule1}). 
Figure \ref{fig2} shows how fast the cascade
grows:
after $g=8$ generations, the unit interval harbours
$N_{8} \sim 3^{8}= 6561$
coefficients on the same number of subintervals.  
Because $r_g = 1$, $\forall g$, 
this fractal cascade is the support for random perturbations to shape the norm of  $\vec{X_g}$ or  
 other dynamical quantities grounded  
on it. The bottom part of figure
\ref{fig2} shows a detail of the fine structure of that object. 
$I_{2}$ is the only segment 
that obeys a self-similar rule, given by 
$ X_{g+1}  = \frac{ X_{g}}{3}$;   $X \in I =   [0,1] $.
Subintervals $I_1$ and $I_0$ don't seem to follow a self-similar
rule. 
The  mapping process shows an independent progression for each of the initial
$I_{i}$ intervals, yielding a non-self-similar structure with
multifractal characteristics whose full  characterization will be carried out elsewhere.
Branch intensities $a_i$, $i = 1 \dots 3^g$,  display a nontrivial behaviour as seen in figure \ref{fig3}.
We plotted $a_i$ for each branch on the limit set after $g=15$ iterations.
The horizontal axis contains $N_{15}=3^{15}=14348907$ points.
Successive enlargements of the initial interval show statistically
equivalent objects as expected for a fractal. 
The figure also shows that the lack of self-similarity extends to both, the branch
positions and their intensities.
The behavior of the distribution of intensities (figure \ref{fig4}) 
is comparable with Thomae's self-similar function as has
been reported also for high-throughput biological
and clinical data \cite{trifonov}.

\section{Cascading scaling behaviour}

\begin{figure}[htb!]
\centering
\includegraphics[width=.9\textwidth]{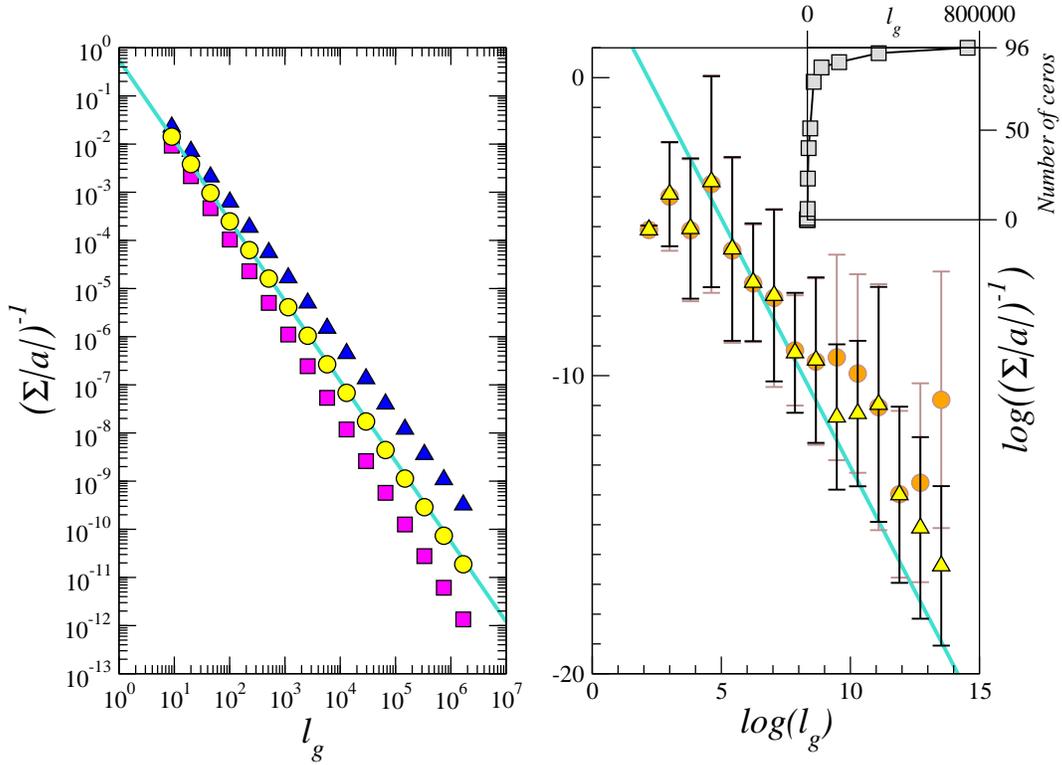}
\caption{ 
In both plots the straight line is a power law with exponent $-\frac{5}{3}$.
Left: Normalized results for $E(l_g) \equiv  \Big( \sum_{\forall a_i \neq 0}  | a_i(   l_g ) |    \Big)^{-1}$.
 Points were calculated from the deterministic backbone with 
  parameter values $a+b=$ (circles) $0.085 + 1.9375 = 2,0225 \sim r_H$, (triangles) $0.085 + 1.786 = 1,871 < r_H$, and (squares) $0.085 +  2.089=2,174 > r_H$). 
  The power law with exponent $-\frac{5}{3}$ is reproduced when the deterministic system is almost tuned on the Hopf bifurcation critical point.
  Right: Results  for $E(l_g) \equiv  \Big( \sum_{\forall a_i \neq 0}  | a_i(   l_g ) |    \Big)^{-1}$, 
  obtained when a uniformly distributed noise, $\eta_g$, is turned on. Parameter values are $r_g=a \eta_g + b$, with $a=1.5$ and $b=2.11$. Circles (averaging over $100$ realizations) are obtained when large deviations at the tails are allowed, while triangles (averaging over $91$ realizations) are for no large deviation allowed (see details in the text).  Error bars represent the relative error calculated as $0.434\frac{\sigma(E)}{E}$, with $\sigma$ the standard deviation.
  Inset: number of obtained zero values as a function of $l_g$.
}\label{fig5new}
\end{figure}

So far, except for using the term cascade to describe successive iterations of rule (\ref{rule0}, \ref{rule1}), 
we haven't shown any link with cascades in turbulence phenomena. 
To establish such a bridge, we must conjecture some sort of parallel between the intensities $a_i$ and the energy by wave number. The idea behind this is that, if $a_i$ is a main changing quantity from the cascading rule  (\ref{rule0}, \ref{rule1}), 
it could be related with a main changing quantity in turbulent cascades: the energy, i.e.,  
\begin{equation}\label{ek}
E(k) \sim E(a_i(g)).
\end{equation}
Thus, we advance the following definition: if there are $l_g$ non-zero intensities at a scale $g$, 
we define an energy measure at generation step $g$ by the relation:
\begin{equation}\label{e2}
E(a_i(l_g)) \equiv  \Big( \sum_{\forall a_i \neq 0}  | a_i(   l_g ) |    \Big)^{-1} ,
\end{equation}
i.e., the inverse value of the summation of the magnitude for all intensities at scale $g$. 
Circles shown in figure  \ref{fig5new}(left) describe the behavior of this quantity as a function of the scale, 
for a deterministic cascade as the one illustrated in figure \ref{fig2}, 
and obtained with $r$ close to  
 the Hopf bifurcation point. 
It can be seen that $E(a_i(l_g))$ obey  a $-5/3$ power law over six orders of magnitude. 
This is the same power law found for the energy spectrum in fully developed isotropic homogeneous turbulence \cite{saddoughi}. 
Therefore,  $E(l_g)$ scales following the expected law for turbulent cascades, 
\begin{equation}\label{e22}
E(l_g) \sim    l_{g}^{-\frac{5}{3}}.
\end{equation}
From this relation, it is straightforward to identify $l_g$ with the wave number $k$ at a given scale, i.e., $l_g \sim k$. 
Thus, assumption (\ref{ek}) can be explicitly written as, 
\begin{equation}\label{ek2}
E(k) \sim E(l_g) \sim    l_g^{-\frac{5}{3}}.
\end{equation}
When the parameter values are distant from the bifurcation critical point,  the calculated $E(l_g)$ deviates 
from the $-\frac{5}{3}$ power law, as is shown in figure  \ref{fig5new}(left). 
The cascading rule  (\ref{rule0}, \ref{rule1})  resembles a turbulent cascade exclusively when the parameter
$r$ is almost tuned on the Hopf bifurcation critical point and such resemblance disappears when the parameter is moved away from it. 
This result emphasizes a critical and deterministic origin of the cascade's $-5/3$ power law. 
Our results suggest that in this deterministic situation, the cascade would continue without end,
meaning that increasing $g$ to a much larger value shall still produce the $-5/3$ law, 
in which case one could assume that the inertial range extends to infinity.
Such an assumption is in line with Onsager's conjecture that
dissipation energy might exist even in the limit of vanishing
viscosity \cite{onsager45,eyink}. Additionally, an interesting characteristic 
of the cascade reported here is that it is rooted in an apparently multifractal structure.

Next, we consider the situation when a uniform noise is turned on. Results for this case are shown in the figure \ref{fig5new}(right).  
We found that, after averaging over realizations, $E(a_i(l_g))$ also follows a $-5/3$ power law, i.e., the analytical cascade behaves on the average as a turbulent system.
Nevertheless, some comments are in order: 
circle points were calculated for $100$ noise realizations. 
For large generations (larger $l_g$) 
the calculated values for $E(a_i(l_g))$ depart from the power law. 
In spite in this region most of the obtained values were zeros,  
a  small number of relatively very large outliers were responsible for the observed departure. 
It is obvious from the inset figure, 
were the number of zeros as a function of $l_g$ is plotted. 
We can see that for the largest $l_g$, $96$ from $100$ outcomes are zeros. 
In this instance only one outlier was enough to cause the departure from the power law. We observe that it is in the large $l_g$ regime where the cascade is more susceptible to the appearance of outliers (i.e., 
more susceptible to uniform randomness) in spite most of the results are dropping to zero.  
So, we also calculated $E(a_i(l_g))$  
excluding $9$ realizations containing at least one outliner for $g \geq 10$.
The obtained result, described by the triangular points at figure \ref{fig5new}(right), fits better to the expected power law.
We interpret the presence of such outliers as indicative of the non adequacy of a uniform noise to mimic realistic fluctuations.
The analysis of a critical tuning of parameters for the noise case is not accessed here. 
However, it is well known that the Hopf bifurcation point is postponed  as a function of the noise intensity \cite{cabreraPhdthesis,cab4}, thus we can't rule out a critical tuning also for the stochastic situation. 
In fact, in light of the deterministic results, it is expected.  

\section{Time delay: a bridge between the DRM and the NS equation}

To the best of our knowledge, a direct relation between the DRM and the NS equation is not currently known. 
Therefore, is up to us to sketch out such a relation. 
So, let's  begin noticing that we can express the solutions, $\vec{u}(\vec{r},t)$, of the  incompressible form of the NS equations in terms of the Fourier representation of $\vec{u}$ as, 
\begin{eqnarray}
\vec{u}(\vec{r},t) &=& \sum_{i} x_i(t) e^{ \vec{k} . \vec{r} } \label{fourier},
\end{eqnarray}
thus resulting in an infinite system of ordinary differential equations for the Fourier coefficients, known as a Galerkin projection of the NS equations \cite{ferrari}, 
\begin{eqnarray}
\dot{x}_i(t)  &=& - \sum_{jk} A_{ijk} x_j(t) x_k(t) - \frac{1}{Re} |\vec{k}|^2 x_i(t) + F_i(t) , 
\end{eqnarray}
where every mode, $x_i$, represents a component of the velocity field with a lengthscale $|\vec{k}|$, the  strength of the nonlinear interaction between modes is given by the coefficients $A_{ijk}$ and $F_i$ is a forcing term. Removing  all but a single arbitrary wave mode, i.e., carrying out a Galerkin truncation, results in 
\begin{eqnarray}
\dot{x}(t)  &=& - A x(t)^2 - \frac{1}{Re} |\vec{k}|^2 x(t) . 
\end{eqnarray}
It is well known that this expression is not formally correct, as a mode cannot interact nonlinearly with itself.  Our results seem to indicate  that a better way to develop the truncation could be by keeping a single mode at two different times, i.e., requiring a time delay, $\tau$, so to obtain  
\begin{eqnarray}
\dot{x}(t)  &=& - A x(t)x(t-\tau) - \frac{1}{Re} |\vec{k}|^2 x(t) . \label{t0}
\end{eqnarray}
For an arbitrary discrete time step, $\Delta$, it can be expressed as 
\begin{eqnarray}
x(t+\Delta)-x(t)  &=& \{- A x(t)x(t-\tau) - \frac{1}{Re} |\vec{k}|^2 x(t) \}\Delta, \label{t1}
\end{eqnarray}
Applying the temporal transformations:
\begin{eqnarray}
\frac{t}{\Delta} &\rightarrow& t^{\prime} \nonumber\\
\frac{t+\Delta}{\Delta} &\rightarrow& t^{\prime} + 1\\
\frac{t-\tau}{\Delta} &\rightarrow& t^{\prime} - h  \nonumber,
\end{eqnarray}
and providing that an integer number $h$ exists, Eq. (\ref{t1}) can be rewritten as, 
\begin{eqnarray}
x(t+1)-x(t)  &=& \{- A x(t)x(t-h) - \frac{1}{Re} |\vec{k}|^2 x(t) \} \label{t2}, 
\end{eqnarray}
where the prime has been omitted. After rearranging terms this equation results in 
\begin{eqnarray}
x(t+1) &=& A x(t) ( \frac{1-Re^{-1} |\vec{k}|^2}{A}- x(t-h)) \label{t3}. 
\end{eqnarray}
Now, requiring $\frac{1-Re^{-1} |\vec{k}|^2}{A} = 1$, we obtain
\begin{eqnarray}
x(t+1) &=& (1-Re^{-1} |\vec{k}|^2) x(t) ( 1- x(t-h)) \label{t4}. 
\end{eqnarray}
That can we expressed as 
\begin{eqnarray}
x_{g+1} &=& r x_g ( 1- x_{g-h}) \label{t5}, 
\end{eqnarray}
where we have defined $r \equiv r_d(h)(1-Re^{-1} |\vec{k}|^2)$, and $t$ has been mapped to the index $g$. $r_d(h)$ is the control parameter value where the difference equation  (\ref{t5}) diverges ($r_d(0)=4$ for the case of the logistic map) so this factor allows us 
to obtain a range of values for $r$ between zero and the divergence value, more information about $r_d(h)$ for different $h$'s can be found in \cite{cabreraPhdthesis,cab1,cab2}.  Eq. (\ref{t5}) is the DRM when $h=1$. It turns out that the inclusion of a time delay at  Eq. (\ref{t0}) is consistent with recent findings of delayed interactions between a range of different spatial and temporal Fourier components \cite{dotti2020,buchhave2021dynamic} and empirically deduced delayed interactions \cite{josserand} in turbulence. In fact, including just one delay may fall short as the delay may increase as one goes to smaller and smaller scales \cite{josserand}.  This observation calls for the inclusion of several delays with changing weights \cite{cabreraPhdthesis} during the cascade, a research that is in progress.

\section{Discussion}

The previous section allows us to establish a more transparent relation between $x$ as described by Eq. (\ref{eq1}) and the cascade coefficients.
Thus, it is worth recalling that eq. (\ref{r33}) 
implies that for a given generation, let's say $g=3$
\begin{eqnarray}\label{dominance2} 
\parallel  \vec{X}_3 \parallel^2 &\sim&  \Lambda_{3} \nonumber\\
x_3^2 + x_2^2 &\sim&  \Lambda_{3} \nonumber\\ 
&=& (a_9+a_8+a_7)r_3^2 +(a_6+a_5+a_4)r_3 + (a_3+a_2+a_1) \label{r34}.
\end{eqnarray}
 That is, the cascade intensity coefficients can be related with the sum of two components of the velocity field with neighbor lengthscales at successive generations. In particular, in the deterministic case ($r=1$)
eq. (\ref{r34}) becomes
\begin{eqnarray}\label{dominance3} 
x_3^2 + x_2^2 &\sim&   \sum_{i=1}^{9} a_i \label{r35}, 
\end{eqnarray}
an expression that can be generalized as
\begin{eqnarray}\label{dominance4} 
x_g^2 + x_{g-1}^2 &\sim&   \sum_{i=1}^{3^{g-1}} a_i \label{r36}.
\end{eqnarray}
or, 
\begin{eqnarray}\label{dominance5} 
x_g &\sim& \sqrt{  \sum_{i=1}^{3^{g-1}} a_i - x_{g-1}^2 } \label{r37},
\end{eqnarray}
i.e., a mode at scale $g$ has information from all the coefficients at scale  $g-1$ (all of them, as $3^{g-1}$ is the total number of coefficients) and from the mode $x_{g-1}$ that itself, carries information from larger scales. This transmission of information through scales is already known from the  rule  (\ref{rule0}, \ref{rule1}) but Eq. (\ref{r37}) convey it in a much transparent way.
Probably, this type of relationship may allows further insights about the cascading process.

Summarizing, here the  $-5/3$ power law was obtained with an exact calculation and not using a numerical approach. It takes us to the question whether turbulence energy cascades could be governed by an analytical structure at least similar to rule  (\ref{rule0}, \ref{rule1}) or whether it may require the inclusion of several delays, in which case 
the effect on the cascading analytical rule and on scaling has to be clarified. Some preliminary work seem to show that a generalization of the rule can be obtained for extended and combined values of the memory $h$.

We would like to finish noticing that these results add further evidence of the relevance of low dimensional discrete dynamics to understand complex phenomena \cite{kadanoff} and turbulence, in particular. 
As intermittency and non-Gaussian statistics 
can be observed in many complex systems 
\cite{cablevy,mantegna}, 
it also may be the case for $-5/3$ turbulent  cascades.
Here it is shown that this cascade is not exclusive for fluids but
a dynamical state rooted solely in critical deterministic nonlinear dynamics. 
In particular, Eq. (\ref{eq1}) is a specific case of a Lotka-Volterra
system,  
known to hold a winnerless competition  \cite{PhysRevE.88.012709} based on a multifractal neural coding scheme, 
which provides the first intrinsic neuro-dynamical explanation for wandering animal behavior \cite{cabSR}. 
Also,  Eq. (\ref{eq1}) can be related to SIR
epidemic transmission dynamics \cite{allen}.
One may conjecture that these two systems could also 
experience complex cascading phenomena of some kind, in nervous information processing in a case and in epidemic propagation in the other.

\section{Acknowledgements}

We thank Juan S. Medina-Álvarez for useful comments and discussions. 
JLCF and EDG acknowledges support from IVIC-141 grant during a small part of this
work and personal support from Prof. M. C. Pereyra (UNM).

\section{Author contributions statement}

JLCF conceived, directed, and developed all aspects of this
research. MRM contributed with software development and simulation
validations, EG 
independently validated the analytics. All
authors contributed to paper writing.
The authors declare no competing financial interests.

\section{Additional information}

Correspondence and requests for materials should be addressed to
JLCF.



\appendix
\section{Eigenvalues of $p_2$}

Let's calculate the second inner product, given by, 
\begin{eqnarray} \label{p2}
p_2 &\equiv&\mathbf{A}_{g-2}^{\dag}  p_1\mathbf{A}_{g-2}= \\ \nonumber
&=&\left[ 
\begin{array}{cc}
\frac{r_{2}}{\beta} & 1 \\ \nonumber
-\alpha r_{2} & 0
\end{array}
\right] \left[ 
\begin{array}{cc}
\frac{r_{1}^2+ \beta ^2}{\beta^2} & -\frac{r_{1}^2}{\beta} \alpha \\ \nonumber
-\frac{r_{1}^2}{\beta}\alpha & \alpha^2r_{1}^2
\end{array}
\right] \left[ 
\begin{array}{cc}
\frac{r_{2}}{\beta} & -\alpha r_{2} \\ 
1 & 0
\end{array}
\right] = \\  \nonumber
&=&\left[ 
\begin{array}{cc}
\frac{%
r_{2}^2r_{1}^2+r_{2}^2\beta^2-2r_{2}r_{1}^2 \alpha \beta^2+r_{1}^2 \alpha^2 \beta^4}{%
\beta^4} & \frac{-r_{2}r_{1}^2-r_{2}\beta^2+r_{1}^2  \alpha \beta^2}{\beta^3} \alpha r_{2}
\\ 
\frac{-r_{2}r_{1}^2-r_{2}\beta^2+r_{1}^2 \alpha \beta^2}{\beta^3}\alpha r_{2} & 
\alpha^2 r_{2}^2 \frac{r_{1}^2 + \beta^2}{\beta^2}
\end{array}
\right] ,  \nonumber
\end{eqnarray}
\noindent with eigenvalues, 
\noindent 
\begin{eqnarray}
\lambda _{\pm}^{p_2} & = & \frac 1{2\beta^4} ( 
r_{2}^2r_{1}^2+r_{2}^2\beta^2-2r_{2}r_{1}^2 \alpha\beta^2+\beta^4 \alpha^2r_{1}^2+\beta^2 \alpha^2r_{2}^2r_{1}^2+\beta^4 \alpha^2r_{2}^2   \pm \nonumber \\
&  &   ( 2r_{2}^4r_{1}^2\beta^2+r_{2}^4r_{1}^4+r_{2}^4\beta^4+2r_{2}^4\beta^6 \alpha^2+\beta^8 \alpha^4r_{1}^4+\beta^8 \alpha^4r_{2}^4-4r_{2}r_{1}^4 \alpha^3\beta^6 \nonumber\\
& & -4r_{2}^3r_{1}^4 \alpha^3\beta^4-4r_{2}^3r_{1}^2 \alpha^3\beta^6+2\beta^6 \alpha^4r_{1}^4r_{2}^2+\beta^4 \alpha^4r_{2}^4r_{1}^4+2\beta^6\alpha^4r_{2}^4r_{1}^2 \nonumber\\
& & -2\beta^8 \alpha^4r_{1}^2r_{2}^2-4r_{2}^3r_{1}^4 \alpha \beta^2+6r_{2}^2r_{1}^4\beta^4 \alpha^2+2r_{2}^4r_{1}^4\beta^2 \alpha^2+4r_{2}^4r_{1}^2\beta^4 \alpha^2 \nonumber\\
& & -4r_{2}^3\beta^4r_{1}^2\alpha+2r_{2}^2\beta^6 \alpha^2r_{1}^2
    )^{1/2}  ). 
\end{eqnarray}

\noindent Here, we can define, 
\begin{eqnarray} 
\Lambda_2 &\equiv&\left[ 
\begin{array}{c}
\left( 1+\beta ^2\alpha ^2\right) { r_1}^2 \\
+\left( \beta^2+\beta ^4\alpha ^2\right) 
\end{array}
\right] { r_2}^2
+\left [ 
\begin{array}{c}
\left( -2\,\alpha\,\beta ^2\right) \,{ r_1}^2
\end{array}
\right] { r_2}
+\left[ 
\begin{array}{c}
\left( \beta^4\alpha ^2\right) { r_1}^2
\end{array}
\right] . \label{A2}
\end{eqnarray}

\noindent It is easy to show that the term under the square root differs from $\Lambda_2$ by  $4\alpha^4r_{2}^2r_{1}^2\beta^8$. 
Then, we can make use of  $\Upsilon_2\equiv \sqrt{4 \alpha^4r_{2}^2r_{1}^2\beta^8}=
2 \beta^4 \alpha^2r_{2}r_{1}
$, to write the eigenvalues of $p_2$ as
$\lambda _{\pm}^{p_2}=\frac 1{2\beta^4}\left( \Lambda_2 \pm \sqrt{\Lambda_2^2-\Upsilon_2^2}%
\right) $. 

\section{Eigenvalues of $p_3$}

The third inner product, given by, 
\begin{eqnarray} 
p_3 &=&\mathbf{A}_{g-3}^{\dag} p_2\mathbf{A}_{g-3}= \\ \nonumber
&=&\mathbf{A}_{g-3}\left[ 
\begin{array}{cc}
\frac{%
r_{2}^2r_{1}^2+r_{2}^2\beta^2-2r_{2}r_{1}^2\alpha \beta^2+r_{1}^2\alpha^2\beta^4}{%
\beta^4} & \frac{-r_{2}r_{1}^2 - r_{2} \beta^2 + r_{1}^2 \alpha \beta^2}{\beta^3}\alpha r_{2}
\\ 
\frac{-r_{2}r_{1}^2-r_{2}\beta^2+r_{1}^2\alpha \beta^2}{\beta^3}\alpha r_{2} & 
\alpha^2 r_{2}^2\frac{r_{1}^2+\beta^2}{\beta^2}
\end{array}
\right] \mathbf{A}_{g-3}, 
\end{eqnarray}
\noindent
has eigenvalues that can be written as 
$\lambda _{\pm}^{p_3}=\frac 1{2\beta^6}\left( \Lambda_3 \pm \sqrt{\Lambda_3^2-\Upsilon_3^2}%
\right) $,  where, 
\begin{eqnarray}
\Lambda_3 &\equiv&\left\{ 
\begin{array}{c}
\left[ 
\begin{array}{c}
\left( 1+\beta ^2\alpha ^2\right) { r_1}^2\\
+\left(\beta ^2+\beta ^4\alpha ^2\right) 
\end{array}
\right] { r_2}^2\\
+\left[ 
\begin{array}{c}
\left(
-2\,\alpha \,\beta ^2-2\,\beta ^4\alpha ^3\right) { r_1}^2
\end{array}
\right] { r_2}\\
+\left[ 
\begin{array}{c}
\left( \beta ^4\alpha ^2+\beta ^6\alpha ^4\right) { r_1}^2
\end{array}
\right] 
\end{array}
\right\} { r_3}^2+ \\ \nonumber
&&+\left\{ 
\begin{array}{c}
\left[ 
\begin{array}{c}
\left( -2\alpha \,\beta ^2\right) \,{ r_1}^2\\
+\left(-2\,\alpha \,\beta ^4\right) 
\end{array}
\right] { r_2}^2\\
+\left[ 
\begin{array}{c}
\left( 2\,\alpha^2\,\beta ^4\right) { r_1}^2
\end{array}
\right] { r_2}
\end{array}
\right\} { r_3}\\
&&+\left\{ 
\begin{array}{c}
\left[ 
\begin{array}{c}
\left( \beta ^4\alpha ^2\right) { r_1}^2\\
+\left( \alpha^2\beta ^6\right) 
\end{array}
\right] { r_2}^2 \nonumber
\end{array}
\right\} \label{A3}
\end{eqnarray}

\noindent and $\Upsilon_3=2\beta^6\alpha^3r_{3}r_{2}r_{1}$.

\section{Eigenvalues of $p_4$}

The fourth inner product becomes, 
\begin{eqnarray}
p_4 &=&\mathbf{A}_{g-4}^{\dag} p_3\mathbf{A}_{g-4}
\end{eqnarray}
\noindent with eigenvalues $\lambda _{\pm}^{p_4}=\frac 1{2\beta^8}\left( \Lambda_4 \pm \sqrt{\Lambda_4^2-\Upsilon_4^2}%
\right) $,  where, 
\begin{eqnarray} 
\Lambda_4 &\equiv&\left( 
\begin{array}{c}
\left\{  
\begin{array}{c}
\left[  
\begin{array}{c}
\left( 1+\,\beta ^2\alpha ^2\right) { r_1}^2\\
+\left(\beta ^2+\,\beta ^4\alpha ^2\right)  
\end{array}
\right] { r_2}^2\\
+\left[  
\begin{array}{c}
\left(
-2\,\beta ^4\alpha ^3-2\,\alpha \,\beta ^2\right) { r_1}^2 
\end{array}
\right] { r_2}\\
+\left[  
\begin{array}{c}
\left( \beta ^6\alpha ^4+\,\beta ^4\alpha ^2\right) { r_1}^2 
\end{array}
\right]  
\end{array}
\right\} { r_3}^2 \\ 
+\left\{  
\begin{array}{c}
\left[  
\begin{array}{c}
\left( -2\,\beta ^4\alpha ^3-2\,\alpha \,\beta ^2\right) { r_1}^2\\
+\left( -2\,\beta ^6\alpha ^3-2\,\alpha \,\beta ^4\right)  
\end{array}
\right] 
{ r_2}^2\\
+\left[  
\begin{array}{c}
\left( 2\,\beta ^6\alpha ^4+2\,\beta ^4\alpha ^2\right) { r_1}^2 
\end{array}
\right] { r_2} 
\end{array}
\right\} { r_3} \\ 
+\left\{  
\begin{array}{c}
\left[  
\begin{array}{c}
\left( \beta ^6\alpha ^4+\,\beta ^4\alpha ^2\right) { r_1}^2\\
+\left( \beta ^8\alpha ^4+\,\alpha ^2\beta ^6\right)  
\end{array}
\right] 
\end{array}
\right\} { r_2}^2
\end{array}
\right) { r_4}^2 \\ \nonumber
&&+\left( 
\begin{array}{c}
\left\{  
\begin{array}{c}
\left[  
\begin{array}{c}
\left( -2\,\alpha \,\beta ^2\right) { r_1}^2\\
+\left(-2\,\alpha \,\beta ^4\right)  
\end{array}
\right] { r_2}^2\\
+\left[  
\begin{array}{c}
\left( 4\,\alpha
^2\,\beta ^4\right) { r_1}^2 
\end{array}
\right] { r_2}\\
+\left[   
\begin{array}{c}
\left( -2\,\alpha^3\beta ^6\right) { r_1}^2 
\end{array}
\right]  
\end{array}
\right\} { r_3}^2 \\ 
+\left\{  
\begin{array}{c}
\left[  
\begin{array}{c}
\left( 2\,\beta ^4\alpha ^2\right) { r_1}^2\\
+\left(
2\,\alpha ^2\beta ^6\right)  
\end{array}
\right] { r_2}^2\\
+\left[  
\begin{array}{c}
\left( -2\,\alpha^3\beta ^6\right) \,{ r_1}^2 
\end{array}
\right] { r_2} 
\end{array}
\right\} { r_3}
\end{array}
\right) { r_4} \\ \nonumber
&&+\left( \left\{  
\begin{array}{c}
\left[  
\begin{array}{c}
\left( \beta ^4\alpha ^2\right) { r_1}^2\\
+\left( \alpha ^2\beta ^6\right)  
\end{array}
\right] { r_2}^2\\
+\left[  
\begin{array}{c}
\left(
-2\,\alpha ^3\beta ^6\right) \,{ r_1}^2 
\end{array}
\right] { r_2}\\
+\left[  
\begin{array}{c}
\left(
\alpha ^4\beta ^8\right) { r_1}^2 
\end{array}
\right]  
\end{array}
\right\} { r_3}^2\right) \label{A4}
\end{eqnarray}

\noindent and $\Upsilon_4=2\beta^8\alpha^4 r_{4}r_{3}r_{2}r_{1}$. 

\section{Eigenvalues of $p_5$}

Here we show $p_5$, given by, 
\begin{eqnarray}
p_5 &=&\mathbf{A}_{g-5}^{\dag} p_4\mathbf{A}_{g-5} ,
\end{eqnarray}

\noindent with eigenvalues, 
$\lambda _{\pm}^{p_5}=\frac 1{2\beta^{10}}\left( \Lambda_5 \pm \sqrt{\Lambda_5^2-\Upsilon_5^2}%
\right) $,  with,

\begin{eqnarray} 
\Lambda_5 &\equiv&\left[ 
\begin{array}{c}
\left( 
\begin{array}{c}
\left\{  
\begin{array}{c}
\left[  
\begin{array}{c}
\left( 1+\beta ^2\alpha ^2\right) { r_1}^2\\
+\left( \beta^2+\beta ^4\alpha ^2\right)  
\end{array}
\right] { r_2}^2\\
+\left[  
\begin{array}{c}
\left( -2\,\alpha\,\beta ^2-2\,\beta ^4\alpha ^3\right) { r_1}^2 
\end{array}
\right] { r_2}\\
+\left[ 
\begin{array}{c}
\left( \beta ^4\alpha ^2+\beta ^6\alpha ^4\right) { r_1}^2 
\end{array}
\right] 
\end{array}
\right\} { r_3}^2 \\ 
+\left\{  
\begin{array}{c}
\left[  
\begin{array}{c}
\left( -2\,\alpha \,\beta ^2-2\,\beta ^4\alpha ^3\right) { r_1}^2\\
+\left( -2\,\beta ^6\alpha ^3-2\,\alpha \,\beta ^4\right)  
\end{array}
\right] 
{ r_2}^2\\
+\left[  
\begin{array}{c}
\left( 2\,\beta ^6\alpha ^4+2\,\beta ^4\alpha ^2\right) { r_1}^2 
\end{array}
\right] { r_2} 
\end{array}
\right\} { r_3} \\ 
+\left\{  
\begin{array}{c}
\left[  
\begin{array}{c}
\left( \beta ^4\alpha ^2+\beta ^6\alpha ^4\right) { r_1}^2\\
+\left( \beta ^8\alpha ^4+\alpha ^2\beta ^6\right)  
\end{array}
\right] { r_2}^2 
\end{array}
\right\}
\end{array}
\right) { r_4}^2 \\ 
+\left( 
\begin{array}{c}
\left\{  
\begin{array}{c}
\left[  
\begin{array}{c}
\left( -2\,\alpha \,\beta ^2-2\,\beta ^4\alpha ^3\right) { r_1}^2\\
+\left( -2\,\beta ^6\alpha ^3-2\,\alpha \,\beta ^4\right)  
\end{array}
\right] { r_2}^2\\
+\left[  
\begin{array}{c}
\left( 4\,\beta ^6\alpha ^4+4\,\beta ^4\alpha ^2\right) { r_1}^2 
\end{array}
\right] { r_2}\\
+\left[  
\begin{array}{c}
\left( -2\,\beta ^6\alpha ^3-2\,\beta^8\alpha ^5\right) { r_1}^2 
\end{array}
\right]  
\end{array}
\right\} { r_3}^2 \\ 
+\left\{  
\begin{array}{c}
\left[  
\begin{array}{c}
\left( 2\,\beta ^6\alpha ^4+2\,\beta ^4\alpha ^2\right) { r_1}^2\\
+\left( 2\,\beta ^8\alpha ^4+2\,\alpha ^2\beta ^6\right)  
\end{array}
\right] { r_2}^2\\
+\left[  
\begin{array}{c}
\left( -2\,\beta ^6\alpha ^3-2\,\beta ^8\alpha ^5\right) { r_1}^2 
\end{array}
\right] { r_2} 
\end{array}
\right\} { r_3}
\end{array}
\right) { r_4} \\ 
+\left( \left\{  
\begin{array}{c}
\left[  
\begin{array}{c}
\left( \beta ^4\alpha ^2+\beta ^6\alpha ^4\right) { r_1}^2\\
+\left( \beta ^8\alpha ^4+\alpha ^2\beta ^6\right)  
\end{array}
\right] { r_2}^2\\
+\left[  
\begin{array}{c}
\left( -2\,\beta ^6\alpha ^3-2\,\beta ^8\alpha ^5\right) { r_1}^2 
\end{array}
\right] { r_2}\\
+\left[  
\begin{array}{c}
\left( \beta ^8\alpha ^4+\beta ^{10}\alpha^6\right) { r_1}^2 
\end{array}
\right]  
\end{array}
\right\} { r_3}^2\right)
\end{array}
  \right] { r_5}^2 \\ \nonumber
&&+\left[ 
\begin{array}{c}
\left( 
\begin{array}{c}
\left\{  
\begin{array}{c}
\left[  
\begin{array}{c}
\left( -2\,\alpha \,\beta ^2\right) { r_1}^2\\
+\left(-2\,\alpha \,\beta ^4\right)  
\end{array}
\right] { r_2}^2\\
+\left[  
\begin{array}{c}
\left( 4\,\alpha
^2\,\beta ^4\right) { r_1}^2 
\end{array}
\right] { r_2}\\
+\left[  
\begin{array}{c}
\left( -2\,\alpha^3\beta ^6\right) { r_1}^2 
\end{array}
\right]  
\end{array}
\right\} { r_3}^2 \\ 
+\left\{  
\begin{array}{c}
\left[  
\begin{array}{c}
\left( 4\,\beta ^4\alpha ^2\right) { r_1}^2\\
+\left(4\,\alpha ^2\beta ^6\right)  
\end{array}
\right] { r_2}^2\\
+\left[  
\begin{array}{c}
\left( -4\,\alpha^3\beta ^6\right) \,{ r_1}^2 
\end{array}
\right] { r_2} 
\end{array}
\right\} { r_3}+\\
\left[ 
\begin{array}{c}
\left( -   2\,\alpha ^3\beta ^6\right) { r_1}^2 \\
\left( -2\,\alpha ^3\beta ^8\right)  
\end{array}
\right] 
{ r_2}^2
\end{array}
\right) { r_4}^2 \\ 
+\left( 
\begin{array}{c}
\left\{  
\begin{array}{c}
\left[  
\begin{array}{c}
\left( 2\,\beta ^4\alpha ^2\right) { r_1}^2\\
+\left(2\,\alpha ^2\beta ^6\right)  
\end{array}
\right] { r_2}^2\\
+\left[  
\begin{array}{c}
\left( -4\,\alpha
^3\beta ^6\right) \,{ r_1}^2 
\end{array}
\right] { r_2}\\
+\left[  
\begin{array}{c}
\left( 2\,\alpha^4\beta ^8\right) { r_1}^2 
\end{array}
\right]  
\end{array}
\right\} { r_3}^2 \\ 
+\left\{  
\begin{array}{c}
\left[  
\begin{array}{c}
\left( -2\,\alpha ^3\beta ^6\right) { r_1}^2\\
+\left(-2\,\alpha^3\beta ^8\right)
\end{array}
\right] { r_2}^2\\
+\left[  
\begin{array}{c}
\left( 2\,\alpha ^4\beta ^8\right) \,{ r_1}^2 
\end{array}
\right] { r_2} 
\end{array}
\right\} { r_3}
\end{array}
\right) { r_4}
\end{array}
\right] { r_5} \\ \nonumber
&&+\left[ \left( 
\begin{array}{c}
\left\{  
\begin{array}{c}
\left[  
\begin{array}{c}
\left( \beta ^4\alpha ^2\right) { r_1}^2\\
+\left( \alpha^2\beta ^6\right)  
\end{array}
\right] { r_2}^2\\
+\left[  
\begin{array}{c}
\left( -2\,\alpha ^3\beta^6\right) \,{ r_1}^2 
\end{array}
\right] { r_2}\\
+\left[  
\begin{array}{c}
\left( \alpha ^4\beta^8\right) { r_1}^2 
\end{array}
\right]  
\end{array}
\right\} { r_3}^2 \\ 
+\left\{  
\begin{array}{c}
\left[  
\begin{array}{c}
\left( -2\,\alpha ^3\beta ^6\right) { r_1}^2\\
+\left(-2\,\alpha ^3\beta ^8\right)  
\end{array}
\right] { r_2}^2\\
+\left[  
\begin{array}{c}
\left( 2\,\alpha^4\beta ^8\right) \,{ r_1}^2 
\end{array}
\right] { r_2} 
\end{array}
\right\} { r_3}\\
+\left\{
 \begin{array}{c}
\left[  
\begin{array}{c}
\left( \alpha ^4\beta ^8\right) { r_1}^2\\
+\left( \alpha ^4\beta^{10}\right)  
\end{array}
\right] { r_2}^2 
\end{array}
\right\}
\end{array}
  \right) { r_4}^2\right]\label{A5}
\end{eqnarray}
\noindent and $\Upsilon_5=2\beta^{10}\alpha^5
r_{5}r_{4}r_{3}r_{2}r_{1}$.

\bibliography{cascadesBIB2}

\end{document}